# Superconductivity at 2.3 K in the misfit compound $(PbSe)_{1.16}(TiSe_2)_2$

N. Giang[1], Q. Xu[2], Y.S. Hor[1], A.J. Williams[1], S.E. Dutton[1], H.W. Zandbergen[2], and R.J. Cava[1]

[1]Department of Chemistry, Princeton University, Princeton NJ 08544 USA

[2]National Centre for HREM, Department of Nanoscience, Delft Institute of Technology, 2628 CJ Delft, The Netherlands

**Abstract**

The structural misfit compound $(PbSe)_{1.16}(TiSe_2)_2$ is reported. It is a superconductor with a $T_c$ of 2.3 K. $(PbSe)_{1.16}(TiSe_2)_2$ derives from a parent compound, $TiSe_2$, which shows a charge density wave transition and no superconductivity. The crystal structure, characterized by high resolution electron microscopy and powder x-ray diffraction, consists of two layers of 1T-$TiSe_2$ alternating with a double layer of (100) PbSe. Transport measurements suggest that the superconductivity is induced by charge transfer from the PbSe layers to the $TiSe_2$ layers.

**Introduction**

Layered $TX_2$ dichalcogenides, where T is an early transition metal and X = S, Se, or Te, exhibit diverse electrical, magnetic, and optical properties [see e.g. 1-3]. The layered dichalcogenides of Nb and Ta show superconductivity with $T_c$s ranging from 0.15 K for 2H-$TaSe_2$ to 7.2 K for 2H-$NbSe_2$, for example. Vacant lattice sites in the van der Waals gap between $TX_2$ slabs can be filled by extra metal atoms, ions, or molecules to alter the magnetic and electronic properties of the parent compounds. $TiSe_2$ has the trigonal symmetry "1T" structure, consisting of layers of edge sharing $TiSe_6$ octahedra stacked along *c*, and has been the topic of research and controversy for decades, as its charge density wave (CDW) transition near 200 K does not fit the conventional picture of electronic instability in two dimensions due to Fermi surface nesting [see, e.g., 5-13]. Normally non-superconducting $TiSe_2$ has recently been made superconducting up to 4 K by Pd and Cu intercalation [14,15].

Misfit compounds, generally described as $(MX)_{1+x}(TX_2)_m$ where M = Sn, Pb, Sb, Bi or a lanthanide; T = Ti, V, Nb, Ta or Cr; X = S or Se; $0.08 < x < 0.28$; and m = 1, 2, 3 [see, e.g. 16-18] are of crystallographic interest due to their unusual structure, which is based on the intercalation of $TX_2$ dichalcogenides with rock salt structure double MX layers; the $TX_2$ and MX structural components have fundamentally different symmetry and periodicity. The inequivalence of the periodicities of the interleaved layers results in a structure that does not match along one in-plane direction, making the crystal structures incommensurate (i.e. "misfit"). The non-integer ratio of MX to $(TX_2)_m$ in the formula is determined by the ratio of the periodicities of the two structural subsystems. Misfit compounds of niobium and tantalum dichalcogenides with rock salt Sn, Pb and Bi monochalcogenide layers have exhibited superconductivity below 6 K [e.g. 19-29]. The $T_c$s are lower than those displayed by the native

$TX_2$ host, except for those based on $TaS_2$, which in pure form has a very low $T_c$ [30]. Here we describe the synthesis and elementary characterization of the misfit compound that results from intercalation of $TiSe_2$ with PbSe layers to form $(PbSe)_{1.16}(TiSe_2)_2$. The compound superconducts at 2.3 K, a case where the intercalation of MX layers induces superconductivity in a normally non-superconducting $TX_2$ host.

**Experiment**

Bulk polycrystalline $(PbSe)_{1.16}(TiSe_2)_2$ suitable for property study was synthesized in a sealed evacuated quartz tube by a method designed to compensate for the vapor transport of PbSe away from the bulk sample under normal synthetic conditions. High purity elements in a 1:1 ratio of PbSe to $TiSe_2$ were sealed in a silica quartz tube under vacuum. The samples were heated first to 350 °C and then at a rate of 50 °C per hour until 650 °C where they were held for twenty hours. The resulting powder was pressed into a pellet and annealed for various times at the optimal misfit synthesis temperature of 900 °C. This resulted in a mixture of the misfit phase plus PbSe. After 16 hours at 900 °C, a succession of two-hour heat treatments at 900 °C was performed to separate excess PbSe from the misfit phase pellet by PbSe vaporization. The samples were quenched in water after every heat treatment. The number of short heat treatments needed to synthesize a pure misfit phase depended on the size of the sample, the size of the silica tubes, and the furnace used. Powder x-ray diffraction (Bruker D8 diffractometer, Cu Kα radiation, diffracted beam monochromator) was used to determine the point at which all PbSe had separated from the sample, leaving behind a single phase misfit compound pellet. X-ray diffraction (XRD) patterns showing no PbSe or $TiSe_2$ peaks are achieved with 20-24 hours total hours of heating; heating beyond 24 hours generally resulted in partial decomposition of the misfit to yield a mixture with $TiSe_2$. To monitor the amount of PbSe lost via vaporization, samples were weighed before and after heat treatments. Pure misfit phase samples were in this way reproducibly found to have PbSe:TiSe$_2$ ratios of 1.0 ± 0.2:2. These ratios are consistent with a misfit compound that contains one double layer of PbSe for every two layers of TiSe$_2$ (see below). Polycrystalline samples of the misfit compound have a dark, silvery appearance. Immediately after heat treatment and quenching, the surfaces of the pellets often had a purple luster due to TiSe$_2$ localization on the surface of the sample. This outer layer was sanded off for sample characterization. Chemical analysis (Galbraith Laboratories) of a single phase sample gave the precise formula (PbSe)$_{1.16}$(TiSe$_2$)$_2$, consistent with the findings from the PbSe weight loss, and in detailed agreement with the structural characterization by electron diffraction (ED) and high resolution electron microscopy (HREM). This is the formula employed in this study. HREM and ED were done with a FEI Titan electron microscope equipped with an aberration corrector and operated at 300 kV.

Misfit crystals were grown using vapor transport, with iodine used as the transport agent. Powder mixtures of Pb, Se, and Ti at the ratio 1(PbSe):1(TiSe$_2$) were heated at 650° C for twenty hours. The powder was then ground and sealed in an evacuated silica tube with a diameter of 15 mm and a length of approximately 20 cm with 90 mg of iodine. The sample was placed in a temperature gradient of 950 – 900 °C, with the powder positioned in the hot end, for eight days. Misfit phase crystals as well as TiSe$_2$ crystals were found in the hot end of the tube. Single crystals ranged from 1 to 5 mm in in-plane dimension and usually were embedded in a polycrystalline mixture of TiSe$_2$, PbSe, and the misfit. The misfit crystals are silver in appearance, in contrast to the purple character of TiSe$_2$, but only through screening the *00l* reflections of crystals by XRD could misfit crystals free of intergrown TiSe$_2$ be selected for study.

The superconductivity of $(PbSe)_{1.16}(TiSe_2)_2$ was characterized through magnetization and resistivity measurements using Quantum Design PPMS and MPMS instruments. The temperature dependent Seebeck coefficient measurements were performed on a home-built apparatus based on MMR Technologies electronics modified to function at low temperatures.

**Results and Discussion**

The TEM images of $(PbSe)_{1.16}(TiSe_2)_2$ show that the structure of the misfit compound consists of alternating double rock salt layers of PbSe and two layers of $TiSe_2$. Figure 1a presents a HRTEM image showing the incommensurate stacking of the two PbSe layers with two $TiSe_2$ blocks. The Fourier transform (Fig. 1b) of the HREM image of which Figure 1a is a part shows two types of diffraction rows along the $c^*$ axis: one with sharp spots and one with diffuse lines. The first one is due to the $TiSe_2$ lattice, which has a long range 3D ordering. The second is due to the PbSe lattice, which has no long range ordering between PbSe double layers along the $c$ axis. The incommensurability of the two lattices is obvious in the HREM image as well as the Fourier transform. Figure 1c shows a HREM images viewed along one of the 'hexagonal' axes of the $TiSe_2$ blocks. The $c$ axis is in the vertical direction. The $TiSe_2$ lattice is clearly visible, but that of the PbSe lattice is not, which is due to the misfit and the fact that the view is not along a crystallographic direction of the PbSe block. The *abc-abc* stacking displayed by the $TiSe_2$ planes in the image is characteristic of the 1T structure, in which the Ti has octahedral coordination with Se. The paired layers of $TiSe_2$ clearly alternate with a double PbSe layer, confirming the misfit as having 1 to 2 ratio of PbSe to $TiSe_2$ plus or minus the incommensurability. The *abc-abc* motif is continued with the same stacking in neighboring $(TiSe_2)_2$ blocks, showing that the PbSe layers do not lead to disorder in the $TiSe_2$ stacking. The angle between the $TiSe_2$ planes and their repeat in the stacking direction seen in Fig. 1a is not exactly 90 ° but 91°, which might be due to an image distortion or a real deviation from 90°, and thus we have not determined definitively whether the compound is orthorhombic or slightly monoclinic.

The precise structural formula of the misfit can be determined by more detailed analysis of the in-plane electron diffraction pattern. The electron diffraction pattern shown in Figure 1d, of the *hk0* reciprocal lattice plane, confirms the presence of two different structural layers, as it shows both square and hexagonal reciprocal lattices. These reciprocal lattices are marked in the figure. The diffraction pattern shows the commensurability of the reciprocal lattices for the -110 hexagonal and 200 cubic reciprocal lattice vectors and their incommensurability in the perpendicular in-plane direction. The two reciprocal lattices become commensurate in this direction after 12 repeats of the $TiSe_2$ layers and 7 repeats of the PbSe layers. Including the fact that the PbSe layers are double, this results in the determination of the structural formula of $(PbSe)_{1.16}(TiSe_2)_2$, in excellent agreement with the formula determined by chemical analysis. Other misfit compounds with this structure type are known: PbS with $NbS_2$ and $TiS_2$, and PbSe with $NbSe_2$ [16,17], for example.

A powder XRD pattern of a polycrystalline sample of the misfit phase is shown in Figure 2. As with many lamellar compounds, a high level of preferred orientation is present. To confirm the structure of the compound, a refinement was done according to previous models of titanium selenide misfits [22], with each layer refined separately. A profile fit that optimizes the intensities of the peaks without taking into account the structural arrangement of the atoms [31] was done for the misfit XRD pattern. Such a fit yielded the best possible agreement for an irregular profile such as is expected for an incommensurate structure. The *c* lattice parameter was determined as 18.247(1) Å from the *00l* reflections, which are marked in the pattern. Fitting of the in-plane cell parameters for the $TiSe_2$ part of the misfit yielded a hexagonal cell parameter of

$a = 3.553(1)$ Å, and fitting of the PbSe part yielded an in-plane tetragonal cell parameter of $a = 6.14(2)$ Å. These values are very close to the cell parameters of TiSe$_2$ ($a = 3.55$ Å) and PbSe (a=6.12 Å) and likely represent a slight relaxation of the cells due to both mutual size accommodation and charge transfer (see below). A schematic of the crystal structure is presented in the inset of Figure 2 – the layers of TiSe$_2$ and the double PbSe rock salt layers are represented, with the stacking repeat of the unit cell defined.

The superconducting transition was characterized by resistivity and susceptibility measurements. Figure 3 shows the in-plane low temperature resistivity in zero field for a (PbSe)$_{1.16}$(TiSe$_2$)$_2$ single crystal. The superconducting critical temperature, at which there is loss of resistivity, is approximately 2.3 K. At higher temperatures, the misfit compound has metallic behavior. The high resistivity ratio, $\rho(300\ \text{K})/\rho(4\ \text{K}) = 18.8$, is indicative of a good metal and suggests that the very irregular bonding between the misfitting PbSe and TiSe$_2$ layers does not strongly scatter the charge carriers. Anisotropy within the plane was not measured. The inset shows a detail of the transition, with normal metallic behavior down to 1.8 K in the presence of a field of $\mu_0H = 0.2$ T perpendicular to the basal plane and the direction of current flow. Rather than completely suppressing the superconductivity in this field, it is likely that T$_c$ has merely been lowered to less than 1.8 K, where it is not detectable in the current measurements. The superconducting transition measured on a polycrystalline sample is broad in the DC susceptibility measurements (inset, Figure 3); the T$_c$ of about 2.3 K is consistent with the resistivity measurements on the single crystal. The broad transition is likely due to a very low value for the lower critical field and the very small crystallite size.

The temperature dependent Seebeck coefficient was measured on a polycrystalline pellet of the misfit compound, and is compared to those for pure and Cu doped TiSe$_2$ [14] in Fig. 4.

For pure TiSe$_2$, the onset of the CDW state is marked by a dramatic change in the Seebeck coefficient near 200 K. This CDW is suppressed with copper doping, yielding an optimal superconductor in Cu$_x$TiSe$_2$ at x = 0.08. At this composition the CDW is no longer present, and the Seebeck coefficient is negative for the full temperature range (Fig. 4), indicating that the Cu intercalation has doped the TiSe$_2$ layer with electrons. Similarly, for the misfit compound (PbSe)$_{1.16}$(TiSe$_2$)$_2$, there is no visible CDW transition, reflecting the suppression of the CDW in TiSe$_2$ by the PbSe intercalation. The negative values of the Seebeck coefficient for the misfit compound indicate that it is n-type for the full temperature range. The extraordinary similarity in Seebeck coefficients for Cu$_{0.08}$TiSe$_2$ and (PbSe)$_{1.16}$(TiSe$_2$)$_2$ suggests a similar electron doping of the TiSe$_2$ layer in the two cases. We therefore speculate that the Seebeck coefficient data indicate that there is charge transfer from the PbSe layer to the TiSe$_2$ layer in (PbSe)$_{1.16}$(TiSe$_2$)$_2$, with the predominant charge carriers in the TiSe$_2$ layers, very similar to what is seen in Cu$_x$TiSe$_2$.

Because both the TiSe$_2$ and PbSe layers are nominally electronically neutral, the reason for the apparent charge transfer between them is not initially clear. The conduction band in TiSe$_2$, which is derived primarily from the Ti 3$d$ orbitals, is very close in energy to its valence band, which is mostly Se 4$p$ in character, resulting in its semiconducting/semimetallic character. PbSe is a small band gap semiconductor with a valence band derived primarily from Se states and a (nominally empty) conduction band derived from Pb 6$p$ states. The inset of Figure 4 shows schematically a general proposal for the TiSe$_2$ and PbSe electronic states in (PbSe)$_{1.16}$(TiSe$_2$)$_2$, represented as a superposition of the density of states from each constituent, a picture that we infer from previous electronic structure models of misfit phases such as (PbS)$_{1.14}$NbS$_2$ and (PbS)$_{1.14}$TiS$_2$ (16). Those models are rigid band like - that is, the only change in the electronic structure of the TX$_2$ host layer upon intercalation of MX layers is a change in band filling of the host. In the current case, charge transfer occurs from the PbSe layer to the TiSe$_2$ layer, resulting in the presence of holes in the PbSe layer and electron doping of the TiSe$_2$ layer. Quantification of the amount of charge transfer is beyond the scope of the current study.

**Conclusions**

Superconductivity is reported at 2.3 K in the misfit compound (PbSe)$_{1.16}$(TiSe$_2$)$_2$. Previously reported superconducting misfits are derivatives of superconducting parent compounds –(PbSe)$_{1.16}$(TiSe$_2$)$_2$ is a rare example of a misfit compound with superconducting behavior derived from charge transfer into a non-superconducting host compound. The parallels between Cu$_x$TiSe$_2$ and (PbSe)$_{1.16}$(TiSe$_2$)$_2$ suggest that superconductivity is induced through a similar mechanism - suppression of the CDW through electron donation to the TiSe$_2$ layer by the intercalants. Electron doping in Cu$_{0.08}$TiSe$_2$ results in conduction that originates from oval-shaped electron pockets at the L-points in the Brillion zone [32]. If the general electronic picture for misfit phases can be applied to (PbSe)$_{1.16}$(TiSe$_2$)$_2$, then in addition to the electron pockets at the L-points there should be hole pockets at other places in the Brillion zone originating from the PbSe layers. Detailed experimental characterization of the electronic structure of (PbSe)$_{1.16}$(TiSe$_2$)$_2$ or other superconducting misfit phases by ARPES, comparing the bands at the Fermi energy to those of the pure TX$_2$ host, would be of significant interest to determine the details of the doping mechanism that results from the MX layer intercalation, and what impact, if any, the incommensurate crystal structure has on the electronic states at the Fermi Energy.

**Acknowledgements**

This work was supported by the US Department of Energy, Division of Basic Energy Sciences, grant DE-FG02-98ER457056.

**Figures**

**Figure 1.** (color on line) TEM images and electron diffraction characterization of the $(PbSe)_{1.16}(TiSe_2)_2$ misfit phase. (a) High resolution image showing the misfit between the PbSe layers and the $TiSe_2$ layers, with the $c$ axis in the vertical direction and the modulation in the horizontal direction. The dark spots image Pb and Se atoms in the PbSe double layers and the Ti and Se atoms in the two $TiSe_2$ layers. The alternating stacking of two $TiSe_2$ layers with double PbSe layers is clearly seen. (b) Fourier transform of the image in Fig. 1a. (c) HREM image viewed along one of the 'hexagonal' axes of the $TiSe_2$ blocks. The $c$ axis is in the vertical direction. (d) The 001 diffraction pattern of the misfit phase. The trigonal reciprocal lattice from the $TiSe_2$ part (red dashed lines) and the tetragonal reciprocal lattice from the PbSe part (black lines) are clearly seen. An orthogonal reciprocal lattice for the $TiSe_2$ part is also shown (red solid lines). Counting the number of repeat units needed to bring the reciprocal lattices (solid black and red lines) to commensurate matching allows determination of the structural formula as $(PbSe)_{1.16}(TiSe_2)_2$, in excellent agreement with the chemical analysis.

**Figure 2.** (color on line) The powder x-ray diffraction pattern for $(PbSe)_{1.16}(TiSe_2)_2$. A profile fit that optimizes positions and intensities of the peaks without refining internal atomic coordinates is shown. The blue points are the experimental data, the red line is the profile fit to the data, and the gray line is the difference between the observed and fit patterns. The blue tick marks (upper) are for the PbSe part and the black tick marks (lower) are for the $TiSe_2$ part. The lattice parameters are: $c = 18.247(2)$ Å, $TiSe_2$ part, hexagonal cell: $a = 3.553(1)$ Å PbSe part, tetragonal cell $a = 6.14(2)$ Å. Inset: schematic of the crystal structure showing the (100) PbSe rock salt double layers and the two 1T-like $TiSe_2$ layers in the crystal structure of $(PbSe)_{1.16}(TiSe_2)_2$.

**Figure 3.** (color on line) Characterization of the superconducting transition and the normal state resistivity for $(PbSe)_{1.16}(TiSe_2)_2$. Main panel: resistivity measurements in the basal plane of a single crystal from 2 to 300 K. Upper inset: detail of the superconducting transition in the resistivity measurements in zero and 0.2 T magnetic field applied perpendicular to the basal plane and the direction of current flow. Lower inset: characterization of the superconducting transition by DC susceptibility measurements at $H_{DC}$ = 3 Oe.

**Figure 4.** (color on line) Seebeck coefficient measurement for $(PbSe)_{1.16}(TiSe_2)_2$. Data for $TiSe_2$ and $Cu_{0.08}TiSe_2$ from reference 14. Insert: Proposed schematic electronic band structure of $(PbSe)_{1.16}(TiSe_2)_2$.

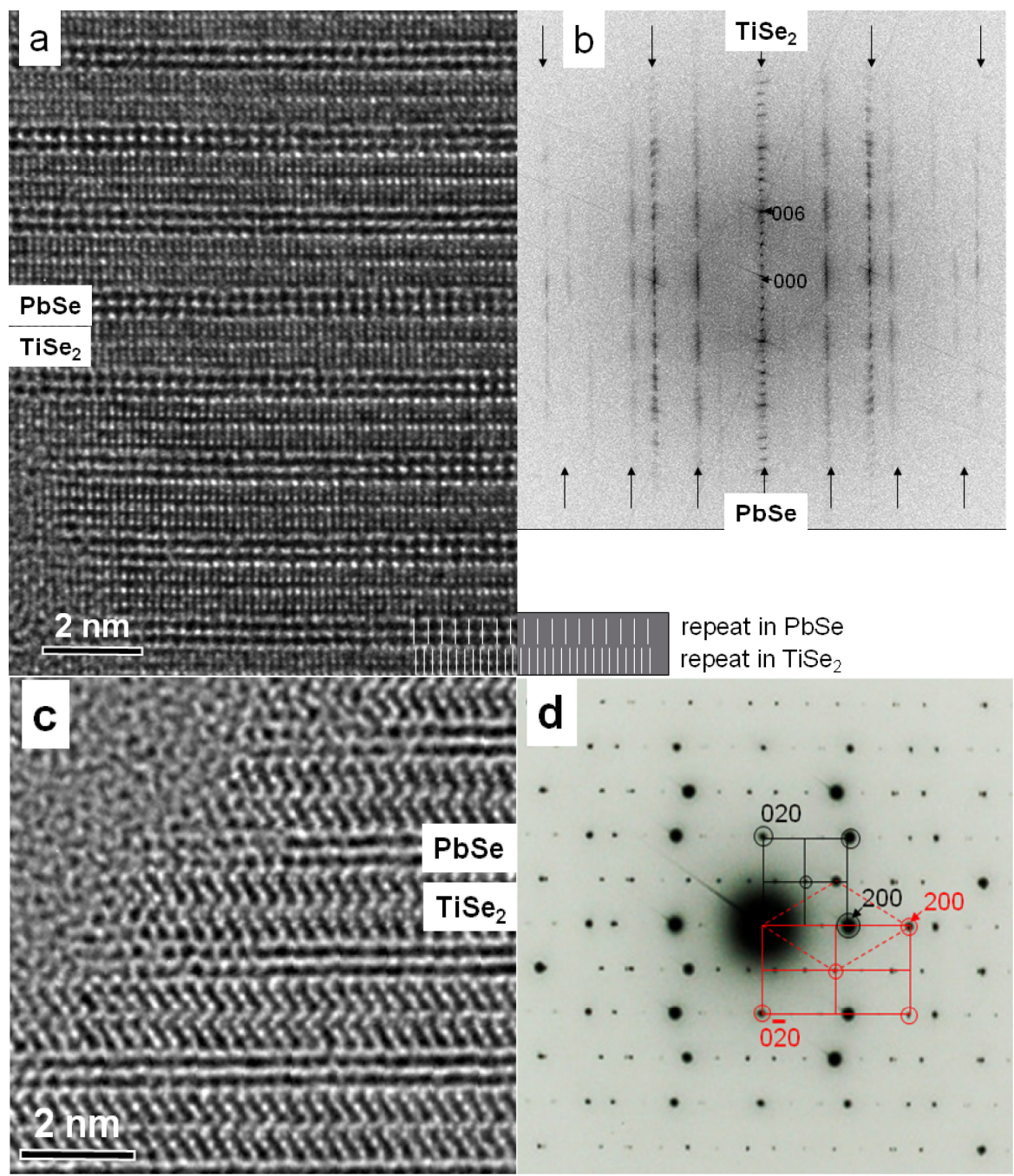

**Figure 1**

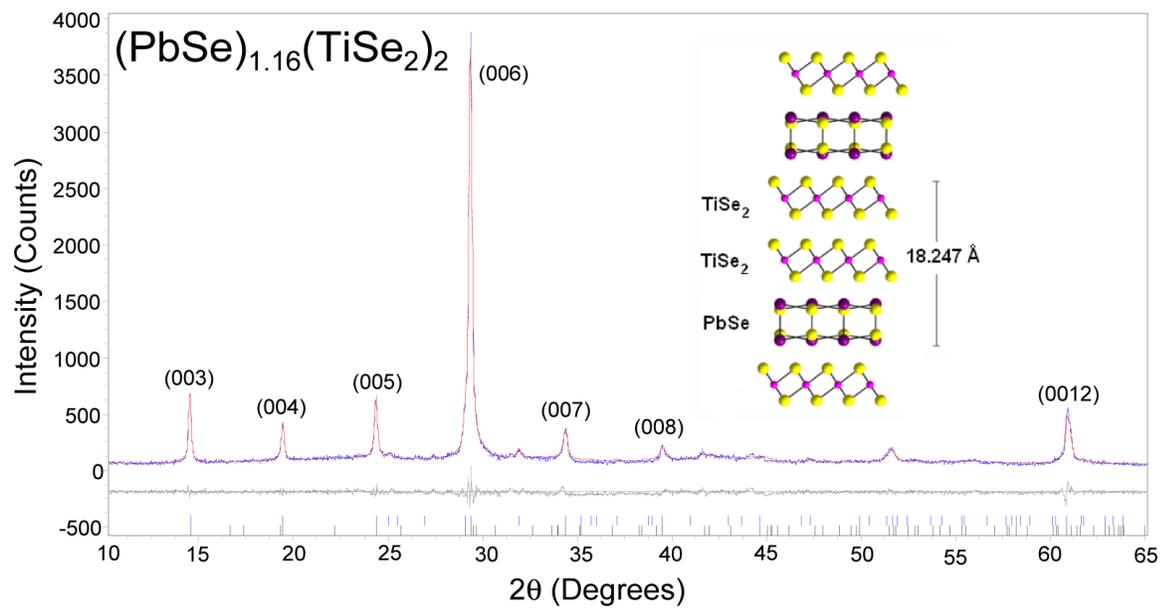

**Figure 2**

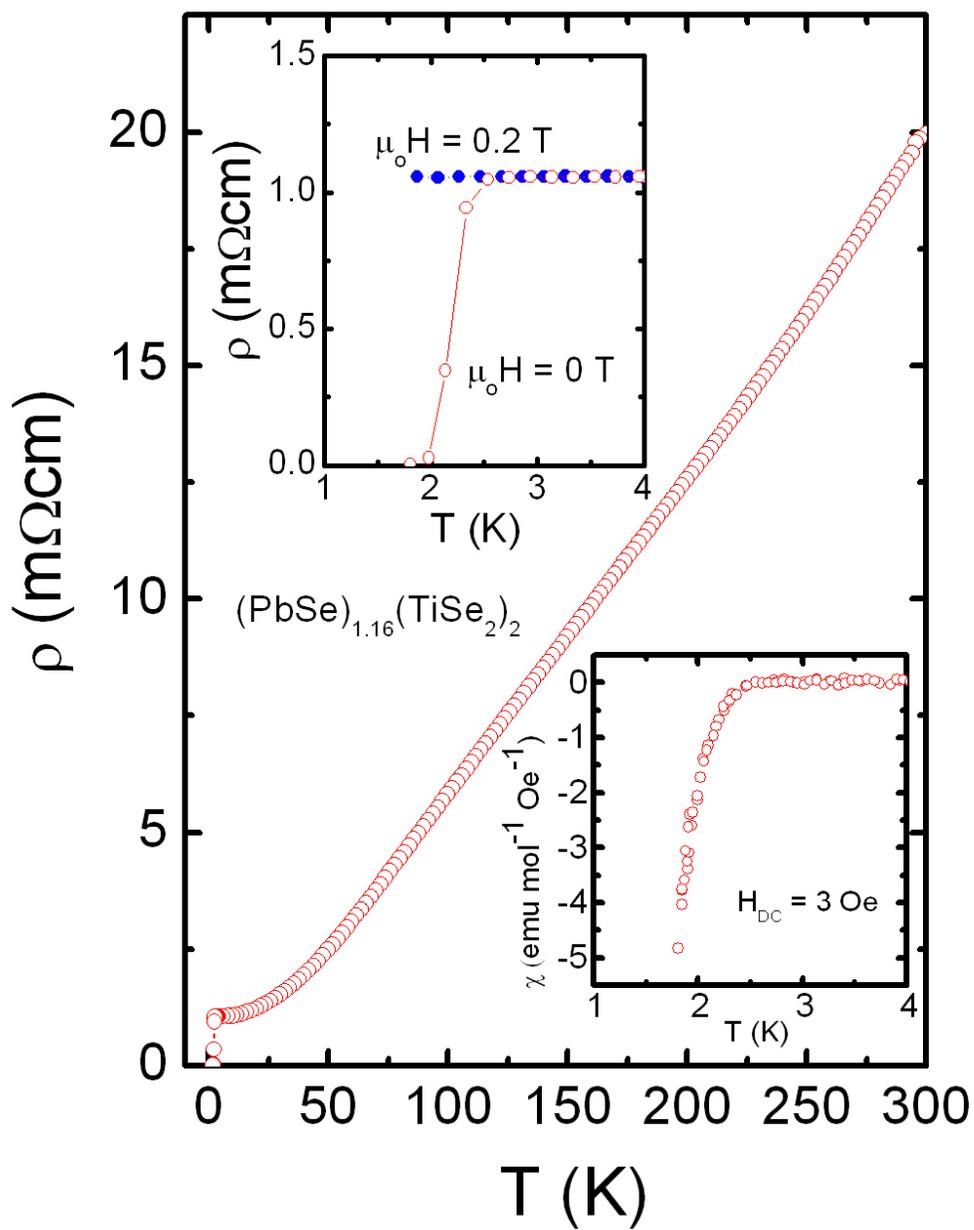

**Figure 3**

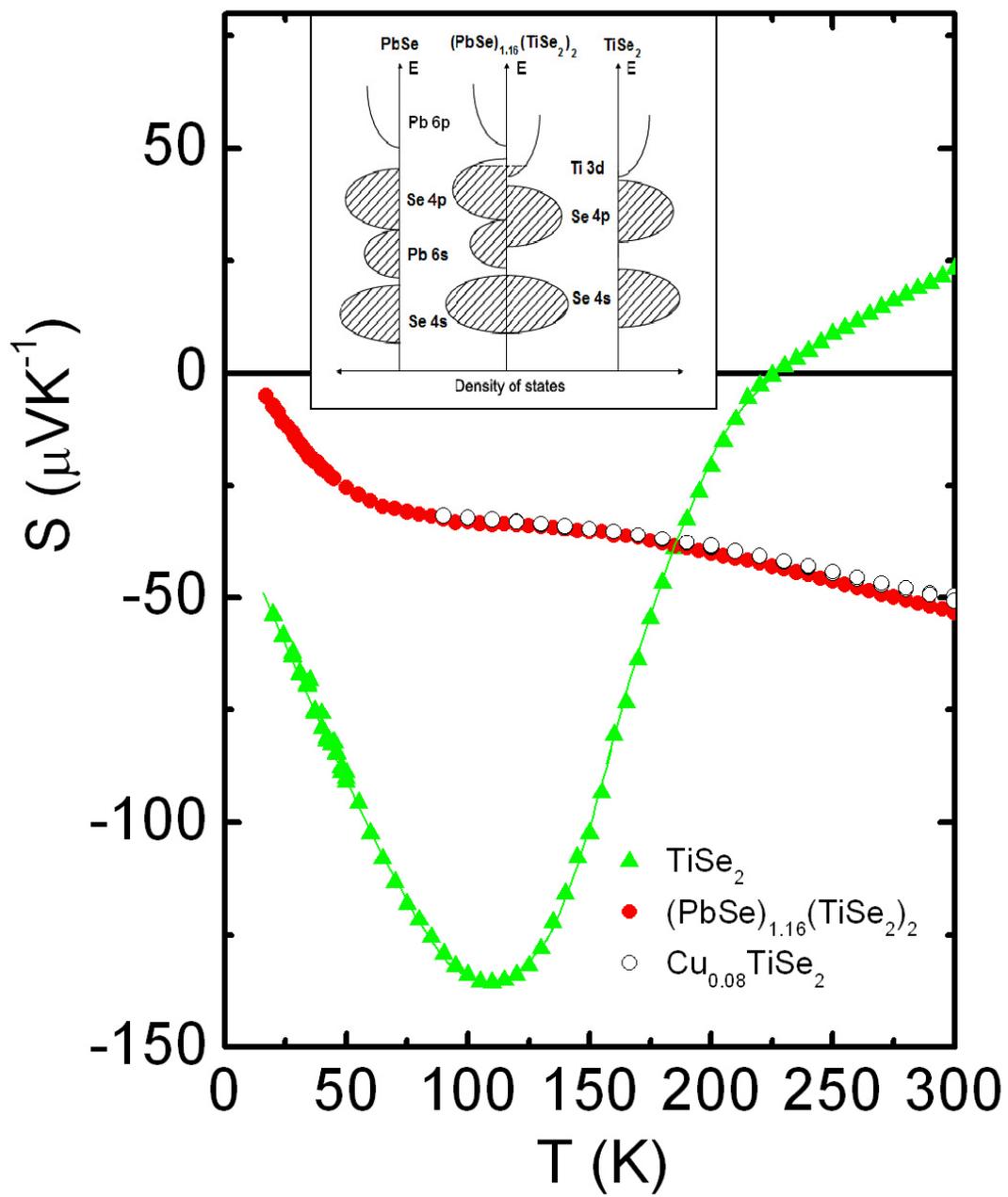

**Figure 4**